\documentclass[conference]{IEEEtran}
\IEEEoverridecommandlockouts
\usepackage{cite}
\usepackage{amsmath,amssymb,amsfonts}
\usepackage{algorithmic}
\usepackage[ruled,vlined]{algorithm2e}
\usepackage{graphicx}
\usepackage{textcomp}
\usepackage{xcolor}
\usepackage{gensymb}
\usepackage{amsthm}
\def\BibTeX{{\rm B\kern-.05em{\sc i\kern-.025em b}\kern-.08em
    T\kern-.1667em\lower.7ex\hbox{E}\kern-.125emX}}

\def\a{\mathbf{a}}
\def\A{\mathbf{A}}
\def\s{\mathbf{s}}
\def\w{\mathbf{w}}

\def\I{\mathbf{I}}

\begin{document}

\title{Phased Array With Improved Beamforming Capability via Use of Double Phase Shifters\\

\thanks{This work was supported by NSF under grant ECCS-2033433}
}

\author{\IEEEauthorblockN{Zhaoyi Xu}
\IEEEauthorblockA{\textit{Department of Electrical and Computer Engineering} \\
\textit{Rutgers University}\\
New Brunswick, USA} \\
\and
\IEEEauthorblockN{Athina P. Petropulu}
\IEEEauthorblockA{\textit{Department of Electrical and Computer Engineering} \\
\textit{Rutgers University}\\
New Brunswick, USA} \\
}

\maketitle

\begin{abstract}
The passive electronically scanned array (PESA) is widely used  due to its simple structure and low cost. {Its antenna weights have unit modulus and thus, only the weights phases  can be controlled. PESA has limited degrees of freedom for beampattern design, where only the direction of the main beam can be controlled.}  In this paper we propose a novel way to improve the beamforming capability of PESA by endowing it with more degrees of freedom via the use of double phase shifters (DPS). By doing so, both the magnitude and the phase of the antenna weights can be controlled, allowing for more flexibility in beampattern design. We also take into account the physical resolution limitation of  phase shifters, and  propose a method to approximate a given complex beamformer using DPS.  Simulation results indicate significant  beamforming improvement  even at low phase resolution.
\end{abstract}


\vspace{-3mm}
\section{Introduction}

Phased array is the most widely used radar.
A passive electronically scanned array (PESA), is a type of phased array,  in which all the antenna elements are connected to a single radio-frequency (RF) chain through phase shifters, which are  controlled by a   computer. The phase shifts are determined so that the  transmitted signal can be focused towards a  desired direction. By changing the phases, the formulated beam can be steered to different directions.
On the other hand, an active electronically scanned array (AESA) is a phased array in which, each antenna element is \textcolor{black}{connected to the RF chain through  a transmit/receive (T/R) module }
under the control of a computer\cite{Kilias2012trmodule}. Each T/R module includes individual power amplifiers, phase shifter and other components. Thus, unlike  PESA, the antenna weights of AESA have controllable magnitude as well as phase, and as a result, can synthesize a more complex beampattern. \textcolor{black}{In a multi-target scenario, AESA can illuminate one target at a time, while creating nulls at other target directions \cite{agrawal2001active}.
} This of course comes at the increased cost of multiple T/R modules, which  limits its application for civilian purposes. AESA are mainly used in military applications \cite{Kilias2012trmodule}. 
Both PESA and AESA transmit a single waveform.



\textcolor{black}{Multiple-input multiple-output (MIMO) radars \cite{Li2008MIMO}  have one RF chain connected to each transmit and receive antenna. The individual analog/digital converters of the RF chain enable MIMO radar to simultaneously transmit multiple waveforms, and  process the signal received by each receive antenna.} When orthogonal waveforms are transmitted, the receive array can use matched filtering to separate the reflection of each    waveform by the target. Those reflections can be used  to formulate a virtual array with much larger aperture than the physical receive array, or equivalently, with much higher  target angle  resolution.
Compared with AESA, MIMO radars can  synthesize multiple distinct complex beampatterns. 
However, the MIMO radar includes digital processing, which involves significant hardware cost and high energy consumption.
Phased-MIMO radar,  a tradeoff between the phased-array and MIMO radars \cite{hassanien2010phasedmimo}, also referred to as   hybrid analog-digital (HAD) beamformer \cite{xu2021learning},   comprises a small number of RF chains  connected to a large number of antennas through a network of phase shifters. The  HAD architecture allows the radar to reap the advantages of MIMO radar with  lower energy cost, however, it incurs a non-convex constant modulus (CM) constraint which complicates the system design.
A way to relax the HAD unit modulus constraint is through the use of double phase shifters (DPS) \cite{Yu2019dps}.

In this paper we propose a novel way to improve the beamforming capability of a PESA by endowing it  with more degrees of freedom. This is achieved via the use of DPS. In particular, each antenna is connected to the sole RF chain via two phase shifters. We will refer to it as DPS-PESA. In  PESA, as the modulus of a phase shifter is $1$, an antenna can  control only the phase of the signal. Therefore, the phases of the antennas can be designed to make all signals add up constructively at the  desired destination. In a multi-target scenario, \textcolor{black}{there is no degrees of freedom to design a beam that can simultaneous focus the energy to one target while reducing the energy to other directions.}
On doubling the phase shifters, one can control both the magnitude and the phase of the waveform transmitted by each antenna without the use of T/R modules. Thus, the use of DPS provides the PESA with improved beampattern design capability, comparable to that of a AESA. 
Compared with AESA, DPS-PESA  involves less costly  hardware, is smaller in size and  consumes much less energy.

In  previous works involving analog phase shifter design \cite{Yu2019dps,Cheng2021dps}, phase shifters with unlimited phase-shift step size has been considered. However, in practical systems, the phase shift step size is a key parameter,  and is typically limited \cite{Yu2011-xg}. 
In this paper, we also take into account the physical resolution limitation of the phase shifters.
However, the discretized phase set makes the design problem an NP-complete linear programming problem. To efficiently solve this problem, we exploit the unique decomposition property of DPS and the normalization of the given beamformer. Finally, a novel algorithm is proposed to approximate given beamforming weights using DPS.


\section{system model and Background}
\label{sec:model}

\begin{figure}[t]
    \centering
    \includegraphics[width = 7cm]{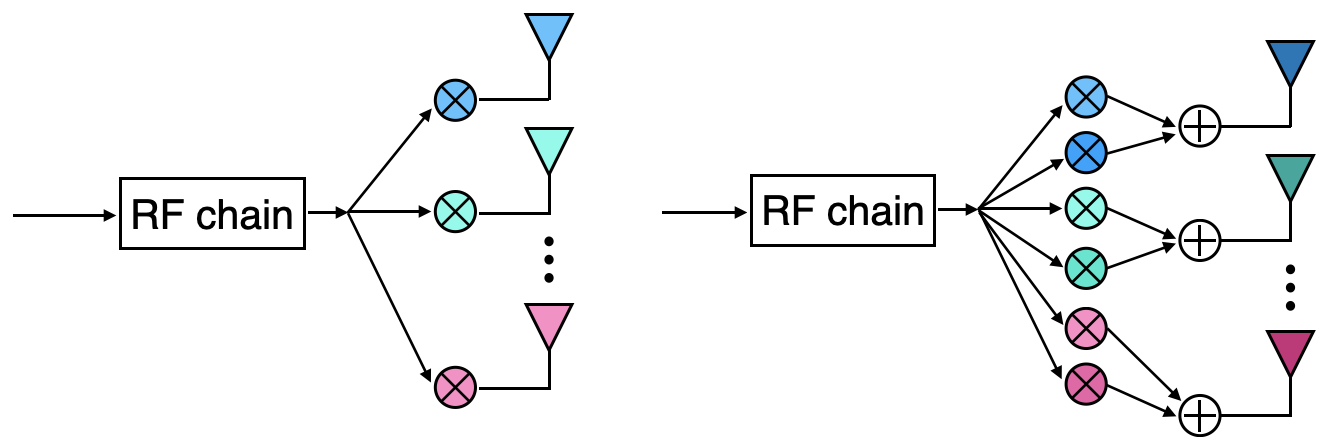}
    \caption{Left: PESA. Right: DPS-PESA}
    \label{fig:phased_array}
    \vspace{-3mm}
\end{figure}

Let us consider a PESA equipped with $N$ antennas and $1$ RF chain. The antennas formulate a uniform linear array (ULA), with spacing between adjacent antennas denoted by $d$.
Each antenna is connected with the RF chain via the phase shifter as shown in the left side of Fig.~\ref{fig:phased_array}.
We assume that the antennas transmit narrow-band signals with carrier wavelength $\lambda$.
The array output at angle $\theta$ is
\begin{equation}
    y(t;\theta) = \a^H(\theta) \s(t),
    \label{sig:y}
\end{equation}
where $\mathbf{a}(\theta) = [1, e^{j 2\pi d\frac{\sin{\theta}}{\lambda}}, \dots, e^{j 2\pi (N-1)d\frac{\sin{\theta}}{\lambda}}]$ is the steering vector at direction $\theta$, and $\s(t) \in \mathbb{C}^{N\times 1}$ is the transmit array snapshot at time $t$. 

For a PESA, the transmitted signal is the baseband signal, $x(t)$, modulated by the antenna phases, i.e.,  
\begin{equation}
    \s(t) = x(t)\w
\end{equation}
where $\w\in\mathbb{C}^{N\times 1}$ is the phase vector and its elements have  constant modulus. Normally, the baseband signal has unit energy.

The PESA output at direction $\theta$ can be written as
\begin{align}
    p(\theta) &= \mathbb{E}\{y(t;\theta) y_s(t;\theta)^H\} \\
    &= \a^H(\theta)\w \w^H  \a(\theta).
    \label{eq:beampattern}
\end{align}
In order to focus the transmitted power towards the desired direction $\theta_0$, we should take $ \w = \a(\theta_0)$. 
Such beamformer is used to track single target since the transmitted power is only maximized at one direction.

\section{Problem formulation }
\label{sec:prob}

\textcolor{black}{Since the phased array only have one analog/digital converter at the receiver side, if multiple targets are illuminated by the radar, the echo signals from them would be coherent and no data-adaptive approach can be used for target detection \cite{Stoica2007probing}.} Thus, in the multi-target scenario, one should only focus on one target while transmitting as less energy as possible to the other targets, ideally, transmitting no energy to those directions. This can be accomplished by null-steering beamformer \cite{Friedlander2012transmit} or the well-known minimum variance distortionless response (MVDR) beamformer \cite{Li2005-qu}. The null-steering beamfomer creates deep nulls at given directions. However, the power to the desired target maybe significantly reduced if any of the nulls is close to the target. On the other hand, MVDR beamformer can mitigate this issue by controlling the depth of the nulls. 

Suppose there are $K$ known targets at different directions $\theta_1, \dots, \theta_K$.
Let the $k$-th target be the target of interest and $\A  = [\a(\theta_1),\dots,\a(\theta_K)]$ contains the steering vectors of all targets.
The corresponding MVDR beamformer can be written as
\begin{equation}
    \w_{m} = (\gamma\I_N + \A\A^H)^{-1} \a(\theta_k)
    \label{mvdr}
\end{equation}
where $\gamma$ controls the depths of nulls. If $\gamma =0$, then the MVDR beamformer \eqref{mvdr} is close to null-steering beamformer. As $\gamma$ increases, the MVDR beamformer \eqref{mvdr} approaches the   beamfomer (see Sec.~\ref{sec:model}).

As one can see, the above beamformer for the multi-target scenario \eqref{mvdr} involves modification on the amplitudes as well as the  phases of the baseband signal transmitted by each antenna. Thus, they cannot be used in PESA to  accomplish for example clutter reduction,  where only the phase of each antenna can be modified. Although such beamformers can be easily realized on AESA, as already mentioned, the cost of AESA is much higher due to the T/R modules.  

Next, we  discuss the use of DPS to improve PESA's beam formulation capability, and approximate the AESA beamforming performance at lower cost.

\section{DPS for PESA}
In DPS-PESA, each antenna is equipped with two phase shifters as shown in the right of Fig.~\ref{fig:phased_array}. The transmit signal is the sum of the phase shifters outputs.
The use of DPS gives PESA the ability to adjust the amplitude and the phase of each antenna like AESA thus a more complex beampattern can be formulated.

Indeed, any complex number with amplitude no larger than $2$ can be uniquely decomposed into two complex numbers with modulus $1$.  Taking complex numbers as vectors, 
\begin{equation}
    a e^{j \omega} = e^{j \phi_1} + e^{j\phi_2}, \quad |a|\leq 2,
    \label{eq:decompose}
\end{equation}
where $\omega = \frac{\phi_1+\phi_2}{2}$ and $a = 2\cos{\frac{\phi_1-\phi_2}{2}}$.
With a given $a$ and $\omega$, one can find the two unit vectors which are symmetric w.r.t. the input vector.


Although we can modify the amplitudes in a PESA like in AESA,  practical phase shifters have a discretized phase control range and usually can not provide the exactly needed phases. Let the phase shifters in a DPS-PESA use $B$-bit representation for the phase,  and have phase control range of $2\pi$. Then, the phase-shift step size is $2\pi/{2^B}$ and the set of all possible phase combinations is denoted as $\mathcal{P}$. The size of $\mathcal{P}$ is $(2^B+1)(2^B)/2$.

Let the beamforming weights of a DPS-PESA with discretized phases be $\Tilde{\w}$.
Our goal is to select those weights so that the beampattern of DPS-PESA approximates as closely as possible the beampattern of an AESA using the same array configuration, i.e., 
\begin{equation}
    \begin{aligned}
        &\min_{\Tilde{\w}} \quad \sum_{i=1}^{M} |\Tilde{p}(\theta_i)-p(\theta_i)| \\
        &\ \textrm{s.t.} \quad \quad \Tilde{w}_n \in \mathcal{P}, \quad n = 1,\dots,N
        \label{question:original}
    \end{aligned}
\end{equation}
where $\Tilde{w}_n$ is the $n$-th element of $\Tilde{\w}$.
Although in \eqref{question:original} the objective function is convex, the optimization problem   falls in the realm of integer programming,  which is NP-complete due to the discrete feasible set. On the other hand, the size of $\mathcal{P}$ grows exponentially with $B$, thus, a  brute-force searching and comparing approach is impractical.

Instead of focusing on approximating the entire beampattern, 
the design problem can be transformed into minimizing the Euclidean distance between $\Tilde{\w}$ and the desired beamformer weights, $\w$. Furthermore, since the decomposition of DPS is unique, we can seek to minimize the error in decomposing each element of $\w$, i.e.,
\begin{equation}
    \begin{aligned}
        &\min_{\Tilde{\w}} \quad ||\Tilde{\w}-\w||_2^2 \\
        &\ \textrm{s.t.} \quad \quad \Tilde{w}_n =  e^{j\phi_a}+ {e}^{j\phi_b}, \\ &\qquad\quad \phi_a, \phi_b \in \mathcal{B}, \quad n = 1,\dots,N
        \label{question:opti}
    \end{aligned}
\end{equation}
where $\mathcal{B} = \{0, 2\pi/{2^B}, \dots, 2\pi(1-1/2^B) \}$ contains all available phases.
To properly approximate the given beamformer, we first decompose each element of the given beamformer into two complex numbers with amplitude $1$ according to \eqref{eq:decompose}. Then, for each complex number we find $L$ closest elements from the available phase set $\mathcal{B}$. After comparing all possible combinations, the one with shortest distance is selected.  The value of $L$ increases with $B$ and is chosen in advance. Since a large $B$ leads to a small step size,  more phases are possible to minimize the total error. 

Since the desired beamformer $\w$ can be scaled by any number as long as the maximum magnitude is no larger than $2$, $\w$ should be properly normalized to minimize the approximation error. 
From \eqref{eq:decompose}, we see that the magnitude of the input vector decides the angle between the two unit vectors. 
If the magnitude of the input vector $a$ is close to $0$, then the difference between $\phi_1$ and $\phi_2$ is approaching $\pi$, and from the gradient of the cosine function we know that a small change of $\phi_1-\phi_2$ would incur a large change of $a$.
On the other hand, if $a$ is close to $2$, then $\phi_1-\phi_2$ is close to $0$, where the cosine function changes slowly. 
Thus before the decomposition, we should normalize the $\w$ to have a maximum magnitude of $2$. 
The whole process is described in Algorithm.~\ref{alg:dps}.

\begin{algorithm}[]
\SetAlgoVlined
\DontPrintSemicolon
Input: $\w$\\
Initialization: Normalize $\w$ to have a maximum magnitude of $2$.\\
\For{$i = 1$ \KwTo $N$}
{Step $1$: Decompose $\w_n  =  e^{j\phi_a}+  e^{j\phi_b}$,  $| e^{j\phi_a}| = | e^{j\phi_b}| = 1$.\\
Step $2$:   
From the available phase set $\mathcal{B}$ find $L$ closest elements to $e^{j\phi_a}$:  $\mathcal{B}_a = \{e^{j\phi_{a1}}, \dots, e^{j\phi_{aL}} \}$ and $L$ closest elements to $e^{j\phi_{b}}$: $\mathcal{B}_b$=$\{e^{j\phi_{b1}}, \dots, e^{j\phi_{bL}} \}$.\\
Step $3$:
From all combinations of $\mathcal{B}_a$ and $\mathcal{B}_b$, find one that is closest to $w_n$ and choose it as $\Tilde{w}_n$.
}{Return;}
\caption{Practical DPS-PESA beamformer design}
\label{alg:dps}
\end{algorithm}

Note that, for phase shifter with unlimited phase-shift step size, connecting each antenna with two phase shifters is enough to provide the special gain, since any beamformer can be normalized to have the maximum amplitude less than $2$ \cite{Yu2019dps}. However, in a  practical setting with limited step size, more than two phase shifters could be connected to one antenna to improve the beampattern approximation performance while the decomposition is no longer unique. 

\section{Numerical Results}
Experiments were conducted to validate the effectiveness of the proposed DPS-PESA by comparing its beampattern with a reference beampattern. The reference beampattern is generated based on \eqref{eq:beampattern} using beamformers in Sec.~\ref{sec:model} and Sec.~\ref{sec:prob}. During the experiments, the given beamformer is approximated by the proposed DPS-PESA following Algorithm.~\ref{alg:dps}.

The first experiment is to  track a single target and the second experiment is single-target tracking with clutter reduction.
Then, to show the significant performance gain of a DPS-PESA, 
we  compare its beampattern performance to that of   PESA using Monte-Carlo experiments in the scenario of tracking single target with clutter reduction.
During the experiments, $N=16$ and $d = 0.5\lambda$ is used and each antenna connects with two phase shifters.



\begin{figure}
    \centering
    \includegraphics[width = 7cm]{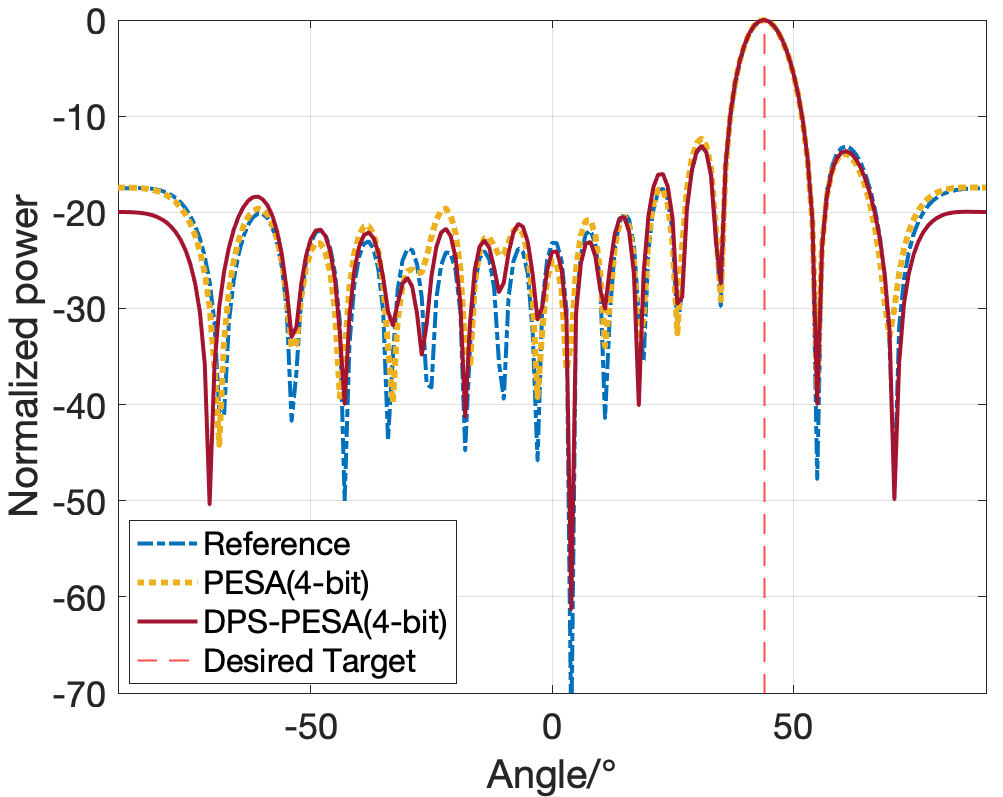}
    \vspace{-3mm}
    \caption{Tracking single target using the proposed DPS-PESA and  PESA.}
    \label{fig:single}
\end{figure}

\textcolor{black}{
In order to show the impact of  limited phase shifter resolution on PESA and the improvement from the use of DPS, we first start with the scenario where a single random target is tracked by the radar. The PESA beamformer is exactly the corresponding steering vector. In this experiment,  $4$-bit phase shifters are used and $L=3$ available phases are used to find the best approximation in Algorithm.~\ref{alg:dps}. For  PESA, the phases of antennas are selected from the phase set $\mathcal{B}$ with minimum difference to the phases of the given beamformer. The corresponding normalized beampatterns are shown in Fig.~\ref{fig:single} where the blue dashed line is the reference beampattern of a PESA without phase resolution limitation, the dotted yellow line denotes the beampattern of  PESA with quantized phases, and the purple solid line represents the beampattern of the proposed DPS-PESA. All the phased arrays can formulate a main beam pointing to the target, while the proposed DPS-PESA enjoys a lower sidelobe level than PESA with quantized phases. The root-mean-square difference between the reference beampattern and PESA is $4.917$dB while that of the proposed DPS-PESA is $4.356$dB.
}

Next, three random targets are generated at $-47\degree$, $30\degree$ and $49\degree$, respectively, and the MVDR beamformer is used to track one of them while suppressing the energy transmitted to the other two targets.
In this experiment, we only want to illuminate the target at $49\degree$ (red dashed line) while the other two are treated as clutter and we want to  to reduce the power delivered to them.
The coefficient $\gamma$ is $0.1$.

\begin{figure}
    \centering
    \vspace{-3mm}
    \includegraphics[width =7 cm]{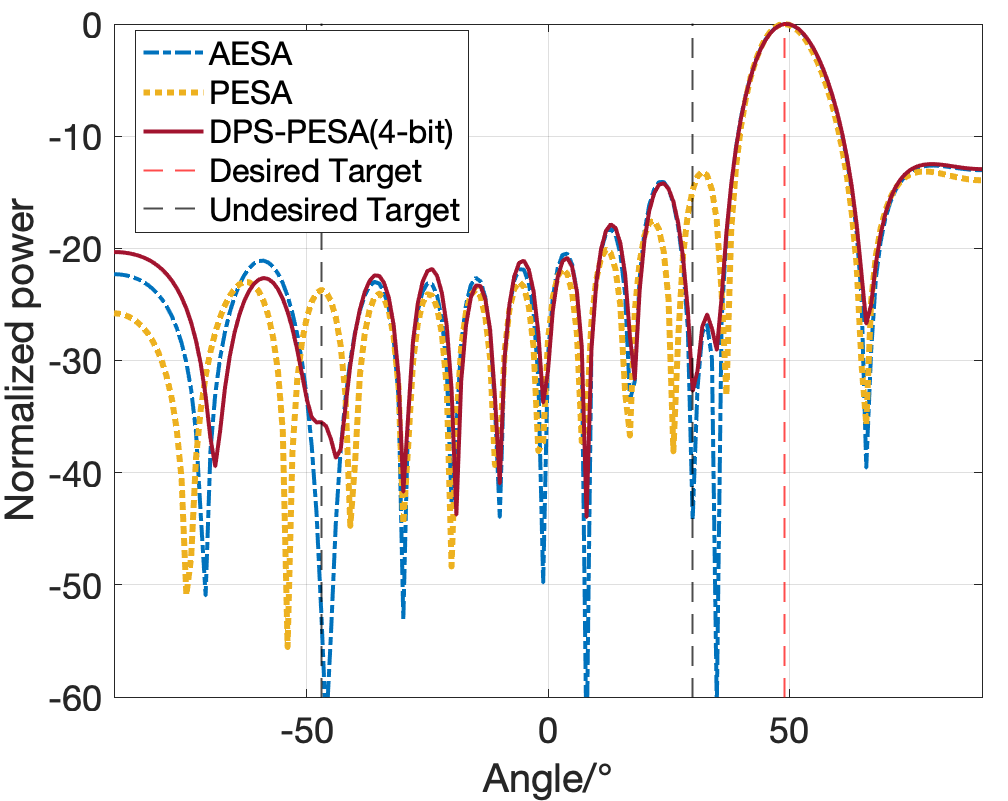}
    \vspace{-3mm}
    \caption{Single target tracking with clutter reduction using MVDR beamformer}
    \label{fig:mvdr}
    \vspace{-4mm}
\end{figure}

Again, $4$-bit phase shifters are used in the proposed DPS-PESA.
 The corresponding beampattern is shown in Fig.~\ref{fig:mvdr}. In comparison, the beampattern of a PESA pointing to $49\degree$ with same array configuration and phase step size is also given in Fig.~\ref{fig:mvdr} while the reference beampattern is from an AESA with unlimited phase step size.
As shown in the figure, the proposed DPS-PESA (purple solid line) has a similar beampattern formulation capability to AESA (blue dashed line);  both methods deliver energy towards the target while suppressing the power level at clutters. 
In  contrast,  PESA (yellow dotted line) can only control the direction of the main beam and can not reduce the power to the undesired targets.  Although the power levels of DPS-PESA are $17.1$dB and $11.4$dB higher than AESA at the undesired targets at $-47\degree$ and $30\degree$ respectively, the beampattern levels of the proposed DPS-PESA are below $-32$dB at those undesired targets which is sufficient for clutter reduction. In comparison,  PESA has a power level higher than $-23$dB at those clutters.

\begin{figure}
    \centering
    \includegraphics[width=7cm]{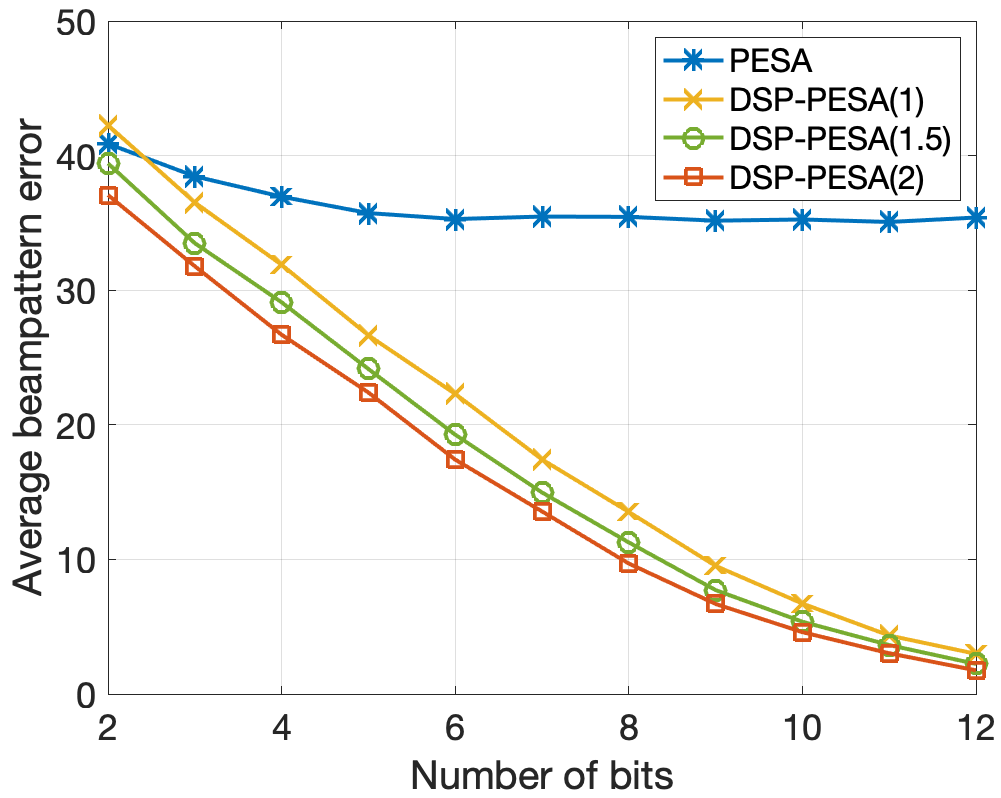}
    \vspace{-3mm}
    \caption{Comparison between PESA and the proposed DPS-PESA.}
    \label{fig:sps}
    \vspace{-3mm}
\end{figure}

In order to quantify the advantages of  DPS and the impact of the number of bits in the phase shifter, we conducted Monte-Carlo experiments to compare the beampattern performance of a PESA and the proposed DPS-PESA.
The proposed DPS-PESA is using a MVDR beamformer with $\gamma = 0.1$.
$2000$ experiments were repeated for each phase shifter bit $B$, and in each experiment three targets are randomly generated with distinct angles and one of them is selected as desired target while the other two are undesired. The reference beampattern is generated based on \eqref{eq:beampattern} using \eqref{mvdr}. The error is the root-mean-square error between the reference beampattern and the beampattern of the proposed DPS-PESA and a PESA at the target and clutter angles. $B$ changes from $2$ to $12$ with a step size of $1$. The result is shown in Figure.~\ref{fig:sps}.
Here the impact of normalization of the desired beamformer is also evaluated. The beamformer is normalized to have a maximum magnitude of $1$ (yellow line), $1.5$ (green line) and $2$ (orange line), respectively. As one can see, when the maximum magnitude is $2$, the overall approximation has minimum error.

From the experiment results, the introduction of DPS significantly benefit the beampattern performance, while maintaining low cost. Although the average beampattern error for the proposed DPS-PESA is above $20$dB when $B=4$, the beampattern of the proposed DPS-PESA is very close to the reference one as shown in Figure.~\ref{fig:mvdr}. The high beampattern error is from the clutter angles since the DPS-PESA cannot create a very deep null due to the limited phase resolution. Note that, when $B=4$, the step size of a phase shifter is $22.5\degree$. On the other side, the increasing of $B$ can reduce the beampattern error. Provided that the error is mainly from the clutter angles, if clutter reduction is highly desired, one should use phase shifters with large $B$ to further reduce the power delivered to the clutter.

\vspace{-2mm}
\section{Conclusions}
We have proposed a novel design of PESA, where each antenna is connected with the  RF chain via two phase shifters. This configuration improves the  beamforming capability of PESA, by adding more degrees of freedom. To solve the integer programming problem that arises in the system design, the unique decomposition of DPS and normalization of beamformer is exploited and a novel algorithm is proposed. Based on numerical results, the use of DPS results in  significant performance improvement.

\bibliographystyle{IEEEtran}
\bibliography{ref}

\end{document}